\documentclass[notitlepage,aps,pra,reprint,onecolumn,longbibliography]{revtex4-1}
\usepackage{graphicx}
\usepackage{amsmath}
\usepackage{amssymb}
\usepackage{comment}
\usepackage[colorlinks, allcolors=blue]{hyperref}
\usepackage[all]{hypcap}
\usepackage[mathlines]{lineno}
\usepackage{braket}
\usepackage{wrapfig}
\usepackage{lipsum}
\usepackage{ulem}

\newcommand{\pref}[2]{\hyperref[#1]{\ref{#1}(#2)}}
\newcommand{\preff}[2]{\hyperref[#1]{\ref{#1}#2}}
\newcommand{\eqpref}[1]{\hyperref[#1]{(\ref{#1})}}

\newcommand{\squig}{{\raise.17ex\hbox{$\scriptstyle\sim$}}}

\newcommand{\tslr}{(t_\text{sys},t_\text{link},t_\text{res})}

\begin{document}
\title{Engineering tunable local loss in a synthetic lattice of momentum states}
\author{Samantha Lapp*}
\author{Jackson Ang'ong'a*}
\author{Fangzhao Alex An*}
\author{Bryce Gadway}
\email{bgadway@illinois.edu}
\affiliation{Department of Physics, University of Illinois at Urbana-Champaign, Urbana, IL 61801-3080, USA}
\date{\today}

\begin{abstract}
Dissipation can serve as a powerful resource for controlling the behavior of open quantum systems.
Recently there has been a surge of interest in the influence of dissipative coupling on large quantum systems and, more specifically, how these processes can influence band topology and phenomena like many-body localization.
Here, we explore the engineering of local, tunable dissipation in so-called synthetic lattices, arrays of quantum states that are parametrically coupled in a fashion analogous to quantum tunneling. 
Considering the specific case of momentum-state lattices, we investigate two distinct mechanisms for engineering controlled loss: one relying on an explicit form of dissipation by spontaneous emission, and another relying on reversible coupling to a large reservoir of unoccupied states.
We experimentally implement the latter and demonstrate the ability to tune the local loss rate over a large range.
The introduction of controlled loss to the synthetic lattice toolbox promises to pave the way for studying the interplay of dissipation with topology, disorder, and interactions.
\end{abstract}
\maketitle

\section{Introduction}

Hermitian Hamiltonians, which describe closed quantum systems, feature unitary time evolution and a spectrum of real energy eigenvalues.
However, real world systems are coupled to their surroundings.
In contrast to closed quantum systems these systems can be described with non-Hermitian Hamiltonians and have eigenvalues which are in general complex.  
The influences of non-Hermiticity, dissipation, and loss on quantum systems have garnered recent interest in several areas.
Researchers have sought to generalize powerful techniques such as optical pumping and dark-state cooling to a many-body context~\cite{Griessner,Diehl-Diss,Tolra-Diss}.
Additionally, there has been interest in how processes like loss and gain can influence and enrich the topological properties of lattice systems~\cite{RudnerLevitov, TonyLee-EdgeExceptional,Leykam-TopNonHerm,Shen-TopNonHermit,Yao-EdgeNonHermit,Gong-Loss-Top,Zeuner-Obs-NonHermi-TopTrans,Zhou-BulkFermiArcs,Zhu-EdgeExceptional-Expt,EdgeSSH} and various types of single-particle and many-body localization phenomena in disordered systems~\cite{AndersonRandom,AndersonLocNon-Hermitian,MBLwLoss-Theory-Levi,MBLwLoss-Theory2-Daley,MBLwLoss-Expt-Luschen}.
The ability to engineer dissipation and artificial environments in cold atom systems has offered a new window into hallmark phenomena associated with quantum electrodynamics~\cite{Oberthaler-Lamb,Krinner-spontanEmiss}.
The use of correlated loss, as well as classical noise~\cite{Stannigel-ClassicalZeno}, has been envisioned as a way of realizing effective constraints on Hamiltonian dynamics for the purpose of stabilizing many-body phases or dynamics of interest~\cite{Zoller-PwaveStabilize}, or for giving rise to unique quantum phases~\cite{Daley-ThreeBody,SafaviNaini-ThreeBody}.
Finally, the detailed study of correlated loss can also be used to probe particle densities or lattice filling factors~\cite{Rempe-Zeno-Mols-2008,yan2013:dipole-dipole,ZhuGadway-ZenoMols-2014} as well as magnetic ordering~\cite{Baur-ProbeOfCorr}.

Some experimental challenges remain, however, when engineering non-Hermitian Hamiltonians. How does one, for example, introduce tunable loss at the level of individual site positions within a lattice without affecting nearby sites? And how does one relate features such as engineered topology and disorder with controlled particle loss?
In this work, we show that so-called synthetic lattices present a natural platform for engineering tight-binding lattice models with controllable local loss, and demonstrate in experiment one method for achieving loss.

In a synthetic lattice~\cite{Boada-SynthDim,Celi-SynthSynth,Mancini2015,Stuhl2015,Wall-Synthetic-Clock,Livi2016,Kolkowitz2017,Sunday-Synth-Mol,Shin-Tubes1,Shin-Tubes2,Chen-Tubes1}, the parametric coupling of discrete quantum states of a particle mimics tunneling along an effective dimension.
By additionally coupling these states to an auxiliary reservoir, loss terms can be introduced organically in synthetic systems for the purpose of studying band topology~\cite{Meier-SSH,Meier-TAI}, disorder~\cite{An-Disorder,An-MobEdge}, and nonlinear atomic interactions~\cite{An-Inter}.
Such capabilities enabled by synthetic lattice systems will complement powerful existing techniques for engineering controlled local loss in real-space lattices~\cite{Ott-LocalizedAtomRemoval-Theory,Ott-ControlledLoss-Expt,Ott-CoherentPerfectAbsorber,Diss-Takahashi,Diss-Sengstock}.

Here, we discuss two mechanisms by which tunable, site-dependent loss can be engineered in synthetic lattices.
Both methods rely on the controlled coupling of the individual sites of a synthetic lattice system to an auxiliary set of quantum states.
The first connects the auxiliary system to a lossy atomic excited state, relying on an explicit form of dissipation by spontaneous emission.
In the second approach, a large reservoir of unoccupied states acts as the auxiliary system, yielding an effective form of loss without explicit dissipation. 
We discuss implementations in the specific context of one-dimensional (1D) synthetic lattices based on linear atomic momentum states; however, these approaches are generalizable to higher dimensions and other experimental platforms.

\section{Engineered loss in synthetic lattices involving multi-level atoms}

In existing synthetic lattice experiments based on laser-coupled linear momentum states~\cite{Meier-AtomOptics,Meier-SSH,An-Disorder,An-FluxLadder,An-Inter,Meier-TAI,An-MobEdge}, all atoms occupy the same hyperfine state, and pairs of Bragg laser beams are used to change the atoms' linear momentum state while leaving the internal state unchanged~\cite{Kozuma-bragg,Stenger-Bragg}.
By including many pairs of Bragg lasers, each addressing a unique two-photon Bragg resonance, many discrete linear momentum states (separated by two photon momenta) can be resonantly coupled to form a synthetic lattice of momentum states.
A natural way to incorporate local loss into this system would be to use momentum-selective Raman-Bragg transitions~\cite{Kas-raman,Weiss-interferom} to change the internal state of population in specific momentum orders (lattice sites), and then remove population from the resulting internal state with resonant light.
We now describe in detail how this mechanism can allow for tunable, site-local loss for atoms in this lattice of momentum states.

%One natural way to incorporate local (in terms of the momentum state index) loss into such a system would be to combine Raman-Bragg transitions~\cite{Kas-raman,Weiss-interferom}, which can change the atomic internal state in a momentum-selective way, with resonant ``removal'' light that is internal state selective but insensitive to the atomic momentum.
%
%We now describe how this mechanismcan allow for tunable, site-local loss within a lattice of momentum states for atoms in the stable internal state.

\begin{figure*}[t]
	\includegraphics[width=\columnwidth]{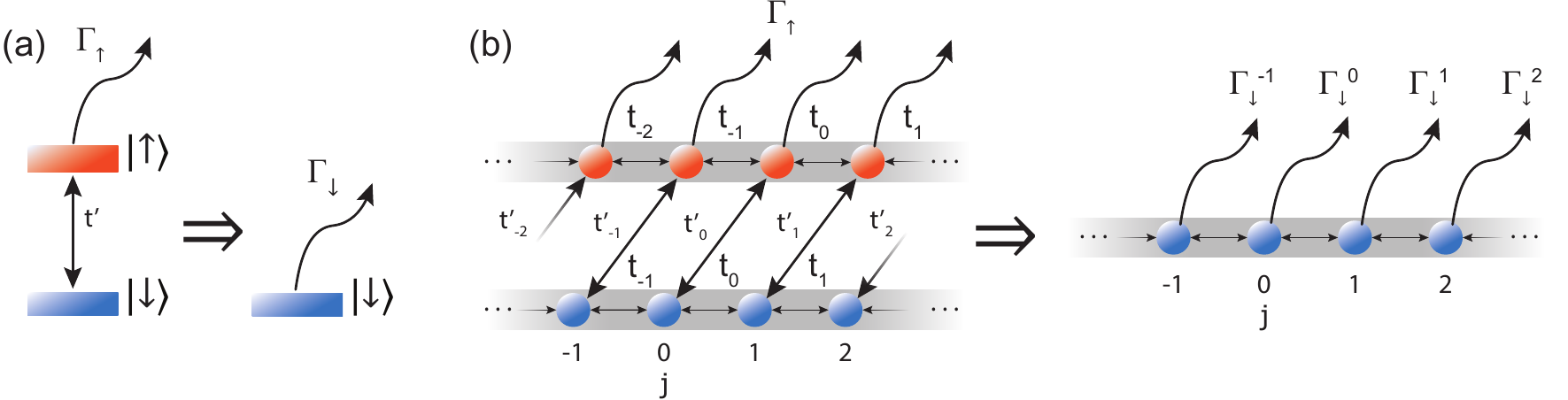}
	\caption{\label{FIG:fig1}
		\textbf{Engineering of tunable, site-dependent loss in synthetic momentum-space lattices.}
		\textbf{(a)}~Engineering effective single site loss using two states, $\ket{\downarrow}$ and $\ket{\uparrow}$.
		A Raman-Bragg transition ($t'$) couples the states and rapid loss in $\ket{\uparrow}$ (achieved by applying a resonant removal laser field) leads to decay of population in $\ket{\downarrow}$.
		This decay of population mimics effective local loss from a single site with loss coefficient given by $\Gamma_{\downarrow} \approx (t')^2 / \Gamma_\uparrow$.
		\textbf{(b)}~Tunable local loss in a many-site synthetic lattice. Bragg transitions connect lattice sites within each internal state, extending the single site loss in (a) to many sites.
		%By driving additional internal-state-preserving Bragg transitions, the single site loss in (a) is easily extended to several sites.
	}
\end{figure*}

We first focus on the simple situation of loss in a system made of just two states coupled with Raman-Bragg laser fields.
%, as shown in Fig.~\pref{FIG:fig1}{a}.
%
We consider the two ground state hyperfine manifolds as they appear for alkali atoms (ground state electron angular momentum $J = S = 1/2$), with hyperfine quantum numbers $F_\pm = I \pm 1/2$ for nuclear spin $I$.
We further restrict the states to have the same magnetic moment, such as the $|F,m_F\rangle = |F_\pm,0\rangle$ clock states for bosonic isotopes.
At low fields, this choice helps to avoid sensitivity of the Raman-Bragg transition to variations of the magnetic field strength.
We explicitly consider a pair of $|F,m_F\rangle$ clock states $\ket{\uparrow} \equiv |2,0\rangle$ and $\ket{\downarrow} \equiv |1,0\rangle$, relevant for species such as $^{23}$Na, $^{39}$K, $^{41}$K, and $^{87}$Rb.
As depicted in Fig.~\pref{FIG:fig1}{a}, we set $\ket{\downarrow}$ as the stable internal state, and $\ket{\uparrow}$ as the ``lossy'' internal state, from which atoms may be effectively removed from the system by applying resonant removal light. This removal manifests as atom loss both from the physical trap as well as from the momentum-space lattice.

%First, ignoring state-preserving Bragg transitions, we focus on how a tunable loss can be created in a system of just two states coupled by Raman-Bragg laser fields. We consider the two ground state hyperfine manifolds as they appear for alkali atoms (ground state electron angular momentum $J = S = 1/2$), with hyperfine quantum numbers $F_\pm = I \pm 1/2$ (for nuclear spin $I$). For practical purposes, we consider coupling two states with the same magnetic moment, such as the $|F,m_F\rangle = |F_\pm,0\rangle$ clock states for bosonic isotopes. At low fields, this choice helps to avoid sensitivity of the Raman-Bragg transition to variations of the magnetic field strength. As depicted in Fig.~\pref{FIG:fig1}{a}, here we explicitly consider a pair of $|F,m_F\rangle$ clock states $\ket{\uparrow} \equiv |2,0\rangle$ and $\ket{\downarrow} \equiv |1,0\rangle$ \rb{[make consistent with figure]}, relevant for species such as $^{23}$Na, $^{39}$K, $^{41}$K, and $^{87}$Rb. We set $\ket{\downarrow}$ as the stable internal state, and the state $\ket{\uparrow}$ as the ``lossy'' internal state, which may be effectively removed from the system of interest by application of resonant removal light (both spatially, and in terms of it's momentum value).

%\rb{make more references to figure 1}

In addition to these two ground hyperfine states, we also implicitly assume that $\ket{\uparrow}$ can be selectively coupled to $\ket{e}$, an excited state, by a one-photon optical transition.
This assumption is valid for all of the alkalis, where the frequency separations of the ground hyperfine manifolds, $E_{\uparrow \downarrow} = E_{\uparrow} - E_{\downarrow}$ (of order hundreds of MHz to several GHz), greatly exceed the excited state loss rates $\Gamma_e$ (of order several MHz for the low-lying excited states accessible via $D_1$ or $D_2$ transitions).
By utilizing optical cycling transitions, many photon momenta may be quickly imparted to atoms in $\ket{\uparrow}$, leading to an effective loss coefficient $\Gamma_\uparrow$ (of order tens to hundreds of kHz), tunable through the intensity, frequency, or stroboscopic control of the cycling light.

We consider atoms initialized in $\ket{\downarrow}$ with roughly zero momentum. By using a pair of counter-propagating Raman-Bragg fields along the $\hat{x}$-axis separated in frequency by $\Delta f = (E_{\uparrow \downarrow} + 4 E^\text{RB}_\text{rec})/h$ (for Planck's constant $h$) the initial state can be coherently coupled to atoms in $\ket{\uparrow}$ and moving with momentum $+ 2\hbar k_\text{RB}$ in the $+\hat{x}$ direction. Here, we assume that the higher frequency field travels in the $+\hat{x}$ direction. 
The recoil energy is given by  $E^\text{RB}_\text{rec} = \hbar^2 k_\text{RB}^2 / 2M$ where $M$ and $\hbar k_\text{RB}$ are the mass and momentum of an atom, respectively.
We further assume that the light fields have roughly equivalent wavelengths $\lambda_\text{RB}$ and wavevectors $k_\text{RB} = 2\pi/\lambda_\text{RB}$, where the wavelength is practically set to a tune-out wavelength between the $D_1$ and $D_2$ lines~\cite{Leblanc-SpeciesSpecific}.
The change in the energy of the light field as a photon is virtually absorbed from one beam and emitted, in a stimulated fashion, into the other beam accounts for the change in energy of the atom in going from $\ket{\downarrow}|p = 0\rangle$ to $\ket{\uparrow}|p = +2\hbar k_\text{RB}\rangle$, i.e. $E_{\uparrow \downarrow} + 4 E^\text{RB}_\text{rec}$.
%
%The coupling strength between these two states will be characterized by an on-resonance two-photon Raman-Bragg Rabi rate, which we simply call $\Omega_\text{RB}$.
%
We let $t'$ represent the two-photon Raman-Bragg coupling strength between these two states.
As we find below, momentum-selectivity of such Raman-Bragg transitions will limit $t'/h$ to be of order ten kHz or less in typical experiments.
Population begins in $\ket{\downarrow}|p = 0\rangle$ with weak Raman-Bragg coupling rate, i.e. $\Gamma_\uparrow \gg t'$.
The rapid loss due to the application of resonant light in $\ket{\uparrow}|p = 2\hbar k_\text{RB}\rangle$  prevents population from coherently building up in this state. In this limit, the dynamics of this system can be effectively mapped to a single-state,  $\ket{\downarrow}|p = 0\rangle$, with an effective loss coefficient $\Gamma_{\downarrow,p=0} \approx (t')^2 / \Gamma_\uparrow$ as shown in Fig. \pref{FIG:fig1}{a}. The scaling of $\Gamma_{\downarrow,p=0}$ with $\Gamma_\uparrow$ and $t'$ reflects the quantum Zeno effect~\cite{Streed-QuantumZeno,Signoles-RydbergZeno}, where enhanced dissipation actually reduces loss in a system by effectively decoupling stable and unstable subspaces. Of great importance to our stated goal of engineering designer loss in a synthetic lattice, this effective loss coefficient for $\ket{\downarrow}|p = 0\rangle$ is tunable through $t'$.

The effective, tunable loss introduced above for the single $\ket{\downarrow}$ momentum state can be extended to a large array of linear momentum states, $\ket{\downarrow}|j\rangle$, by driving a set of two-photon Bragg transitions.
We note that these transitions differ from the Raman-Bragg transitions by both preserving the internal state and by imparting different amounts of momentum.
These Bragg transitions connect linear momentum states $p_j=2j\hbar k_\text{B}$, quantized in units of the photon recoil momentum $2 \hbar k_\text{B}$, where $k_\text{B}$ is the wavevector of the Bragg fields.
Associated with this Bragg wavevector is the Bragg recoil energy $E^\text{B}_\text{rec} = \hbar^2 k_\text{B}^2/2M$.

Raman-Bragg transitions can then couple atoms between the two spin states in a momentum-dependent fashion: for transitions from an initial state $\ket{\downarrow}\ket{j}$ with momentum $p_j = 2 j \hbar k_\text{B}$ to a final state $\ket{\uparrow}\ket{j}$ with momentum $p_j + 2\hbar k_\text{RB}$, the resonant Raman-Bragg condition involves an energy change of $E_{\uparrow \downarrow} + (2\hbar^2/M)(2 j k_\text{B} k_\text{RB} + k^2_\text{RB})$.
The explicit $j$-dependence of this Raman-Bragg resonance condition allows for a local, momentum- or site-dependent coupling of $\ket{\downarrow}$ atoms to the lossy state, $\ket{\uparrow}$.
To ensure momentum selectivity, the rate of individual Raman-Bragg transition couplings should be less than ten kHz for typical conditions, i.e. much less than $4\sqrt{E^\text{B}_\text{rec}E^\text{RB}_\text{rec}}$ to avoid off-resonant driving. By simultaneously driving many Raman-Bragg transitions, many sites can independently be coupled to the ``lossy'' manifold, $\ket{\uparrow}$, at different rates $t'_{j}$. Assuming momentum-independent loss at a rate $\Gamma_\uparrow$ from $\ket{\uparrow}$, this allows for an effectively tunable, site-dependent loss in the synthetic lattice of momentum states, $\ket{\downarrow}|j\rangle$.

The engineering of controllable local dissipation can be combined naturally with the ability to engineer an effective ``tunneling'' between the $\ket{\downarrow}|j\rangle$ sites as shown in Fig. \pref{FIG:fig1}{b}. As long as all of the individual Raman-Bragg coupling rates $t'_{j}$
%(relating to coupling from the the state $\ket{\downarrow}|j\rangle$ to the associated $\ket{\downarrow}$ state with momentum $2 j \hbar k_\text{B}  + 2\hbar k_\text{RB}$)
are much lower than the loss coefficient $\Gamma_\uparrow$, each site will experience a tunable loss coefficient $\Gamma_{\downarrow,j} \approx (t'_{j})^2 / \Gamma_\uparrow$. This scheme can easily be implemented in the context of momentum-space lattices~\cite{Gadway-KSPACE,Meier-AtomOptics}.

We now compare the dynamics under two lossy Hamiltonians: one ``full'' version that features a loss term $\Gamma_\uparrow$ acting only on $\ket{\uparrow}$, and an ``effective'' Hamiltonian that includes only $\ket{\downarrow}$, but with effective, site-dependent loss rates $\Gamma_{\downarrow , j}$. The ``full'' model, which implements loss by exposing $\ket{\uparrow}$ to cycling light, is given by
\begin{equation}
H_\text{full} = - \sum_{\sigma} \sum_{j}\left(t_j \hat{c}^{\dagger}_{\sigma , j+1}\hat{c}_{\sigma, j} + \text{h.c.}\right) - \sum_{j}\left( t'_{j}\hat{c}^{\dagger}_{\uparrow,j} \hat{c}_{\downarrow,j}+\text{h.c.}\right) + i \Gamma_\uparrow \sum_{j} \hat{c}^{\dagger}_{\uparrow, j}\hat{c}_{\uparrow,j} \ .
\label{EQ:FullHam}
\end{equation}
Here, the $\hat{c}_{\sigma , j}^\dagger$ ($\hat{c}_{\sigma , j}$) terms create (annihilate) a particle in the internal state $\sigma$ ($\ket{\downarrow}$ or $\ket{\uparrow}$). Implicitly, the index $j$ relates to linear momentum states with $p = 2 j \hbar k_\text{B}$ along the $\hat{x}$ direction for $\ket{\downarrow}$ and momentum  $p = 2 j \hbar k_\text{B} + 2 \hbar k_\text{RB}$ for $\ket{\uparrow}$. Without loss of generality, we also assume all Bragg ($t_j$) and Raman-Bragg ($t'_{j}$) ``tunneling'' terms to be purely real.

\begin{figure*}[t]
	\includegraphics[width=\columnwidth]{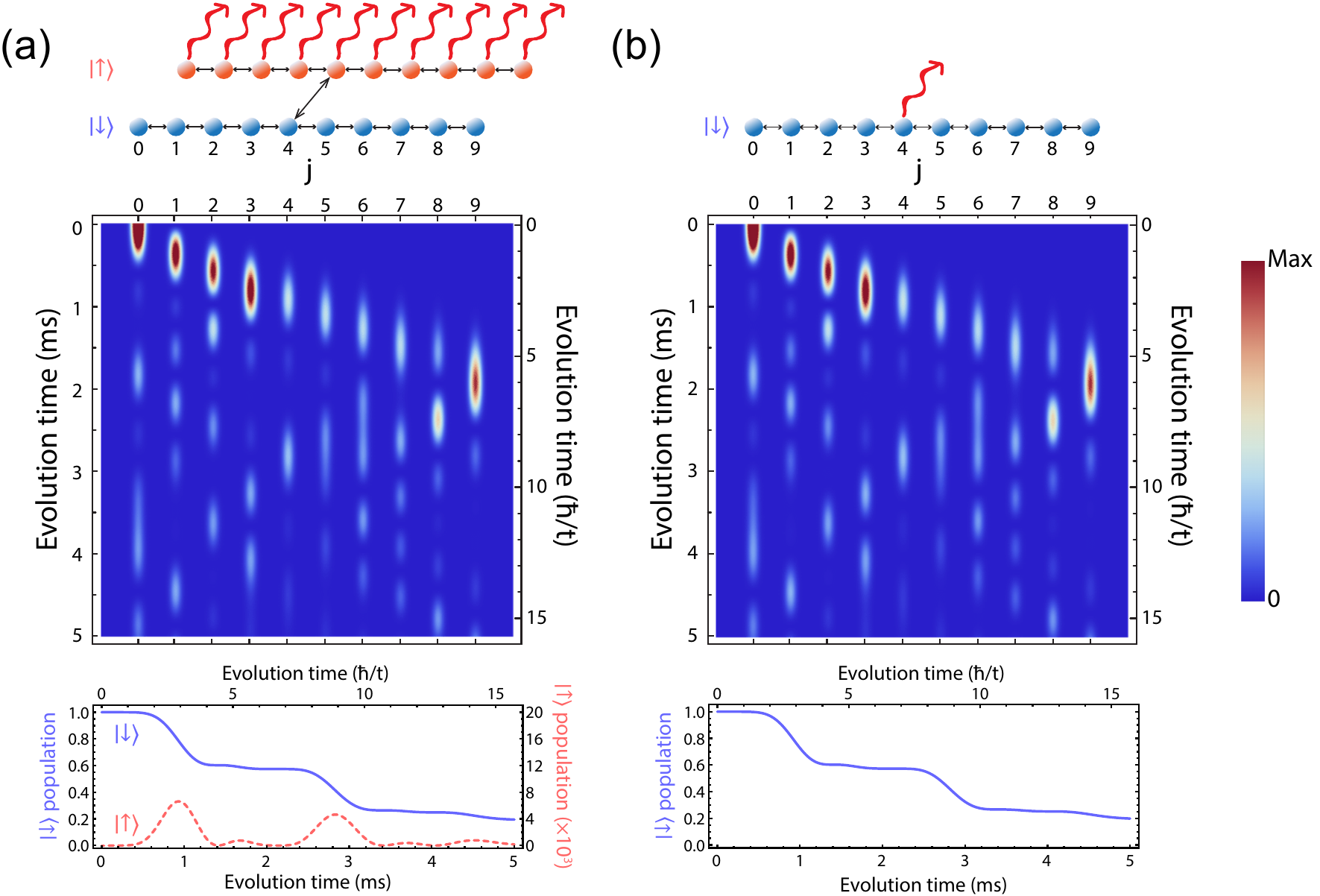}
	\caption{\label{FIG:fig2}
		\textbf{Partial reflection from a lossy ``barrier'' in a synthetic lattice.}
	   \textbf{(a)}~A lossy barrier implemented by coupling to an auxiliary set of lossy states (``full'' model).
	   Top, depiction of the model, with red arrows showing loss out of $\ket{\uparrow}$. Both 1D lattices ($\ket{\uparrow}$ and $\ket{\downarrow}$) have uniform tunneling $t/h = 0.5$~kHz, and the lossy barrier at site $j=4$ connects the two spin states with tunneling $t'/h = 2$~kHz. The $\ket{\uparrow}$ sites feature a global uniform loss of $\Gamma_{\uparrow}/h=10$~kHz.
	   Middle, numerical simulation of the evolution of population initialized at site $j=0$ on the $\ket{\downarrow}$ lattice.
	   Bottom, total population in $\ket{\downarrow}$ (solid blue curve) and $\ket{\uparrow}$ (dashed red curve). Jumps in population coincide with population hitting the lossy barrier at site 4.
       \textbf{(b)}~The effective model of a lossy barrier with a direct loss term.
       Top, depiction of the effective model, with red arrow showing local loss at site 4, $\Gamma_{\downarrow, 4}/h=0.4$~kHz. Tunneling rates are again $t/h=0.5$~kHz.
       Middle, numerical simulation of the evolution of population initialized at site $j=0$ on the 1D lattice.
       Bottom, total population in the 1D lattice as a function of time. Jumps in population again coincide with population hitting the lossy barrier at site 4.
	}
\end{figure*}

The effective model that describes the dynamics purely in $\ket{\downarrow}$, is given by
\begin{equation}
H_\text{eff} = - \sum_{j}\left(t_j \hat{c}^{\dagger}_{j+1}\hat{c}_j + \text{h.c.}\right) + i \sum_{j} \Gamma_{j} \hat{c}^{\dagger}_{j}\hat{c}_{j} \ .
\label{EQ:FakeHam}
\end{equation}
Here, the $\Gamma_j$ terms are equivalent to the $\Gamma_{\downarrow , j}$ terms described above, as the description is purely in terms of $\ket{\downarrow}$ atoms.

In Fig.~\ref{FIG:fig2} we directly compare the dynamics under the full model Eq.~\eqref{EQ:FullHam} with those under the effective model Eq.~\eqref{EQ:FakeHam}.
These should be equivalent when beginning in $\ket{\downarrow}$ and for cases with $t'_{j} \ll \Gamma_{\uparrow}$ for all $j$.
We numerically simulate a situation in which an effective loss appears only at one site of a synthetic lattice, and investigate the dynamics that result from the case of population initially localized at the leftmost site (beginning purely in $\ket{\downarrow}$ for Eq.~\eqref{EQ:FullHam}).

For the full model we consider two 1D synthetic lattices each consisting of 10 sites with nearest-neighbor coupling $t/h=0.5$~kHz as shown in Fig.~\pref{FIG:fig2}{a, top}.~The two lattices, representing $\ket{\uparrow}$ and $\ket{\downarrow}$, respectively, are coupled at site 4 via the Raman-Bragg coupling scheme where $t'_{4}/h=2$~kHz. Additionally, global uniform loss is included in $\ket{\uparrow}$ with strength $\Gamma_{\uparrow}/h=10$~kHz.

Population is initially localized on site $j=0$ and then allowed to evolve. As shown in Fig.~\pref{FIG:fig2}{a, middle}, population initially coherently transfers out of site 0 but is partially reflected at site 4. The transmitted fraction of the population continues until it encounters the edge of the lattice at site 9 and reflects back. Unlike reflection at site 9, every time population reflects from site 4, the total population in $\ket{\downarrow}$ drastically decreases leading to a step-like profile as shown by the solid blue curve in Fig.~\pref{FIG:fig2}{a, bottom}. This population reduction is due to the fact that reflection at site 4 is also accompanied by transfer of population to $\ket{\uparrow}$. The transferred population is, however, quickly lost from the system due to the strong global loss in $\ket{\uparrow}$. The total population in $\ket{\uparrow}$ therefore briefly builds up whenever there is reflection at site 4 as shown by the dashed red curve.

In the regime where $t'_{j} \ll \Gamma_{\uparrow}$ we expect the full model to map onto an effective model with an effective local loss at site 4 given by $\Gamma_{\downarrow,4} \approx (t'_{4})^2 / \Gamma_\uparrow$. For this effective model we consider a 1D lattice consisting of 10 sites as shown in Fig.~\pref{FIG:fig2}{b, top}. The nearest-neighbor tunneling is given by $t/h=0.5$~kHz while the effective local loss is given by $\Gamma_{\downarrow,4}/h=0.4$~kHz. Population is initially localized at site 0 and is then allowed to evolve as shown in Fig.~\pref{FIG:fig2}{b, middle}.
The dynamics agree nearly identically with the full model, showing parts of the population reflecting from and transmitting through the lossy ``barrier'' at site 4.
%In agreement with the case of the full model, population initially coherently moves out of site 0 towards site 9, but is partially reflected at the lossy ``barrier'' at site 4. The fraction of the population transmitted past site 4 continues to move until it encounters the edge of the system and is reflected back into the system.
%
As shown in Fig.~\pref{FIG:fig2}{b, bottom} the total population in the lattice drastically decays whenever population is reflected at site 4, leading to a step-like profile akin to the case of the full model.
In the $t'_{j} \ll \Gamma_{\uparrow}$ limit, we have found good agreement between the full-model and effective-model simulations of Fig.~\ref{FIG:fig2}, confirming the protocol for implementing local loss through coupling to an auxiliary, lossy set of states.

We note that while we have only described how dissipation may be engineered into 1D synthetic lattices with nearest-neighbor tunneling terms, this scheme naturally extends to situations with longer-range hopping terms~\cite{An-MobEdge} or higher-dimensional lattices~\cite{Gadway-KSPACE,An-FluxLadder}.

\section{``Loss'' without dissipation: reversible coupling to a large reservoir of states}

In the previous scenario, we invoked a natural form of dissipation from atomic physics experiments - spontaneous emission - to create a controlled, effective loss in a synthetic lattice.
This scheme involved two key elements: first, we assumed that the states in the ``lossy'' subspace ($\ket{\uparrow}\ket{j}$) could be strongly coupled to a near continuum of states (many different final momentum values after multiple absorption-spontaneous emission cycling events), such that the probability of returning to the initial state was essentially zero.
Second, we assumed that the coupling rate between the stable and lossy subspaces was much smaller than the spontaneous emission loss rate from the lossy states.
This assumption ensured that population would not coherently build up in the lossy subspace, but would instead be lost from the system.

%In the above scenario, we invoked a natural form of dissipation appearing in atomic physics experiment - spontaneous emission - to describe how one may engineer a controlled, effective loss in synthetic lattices. This scheme involved two key elements. First, we assumed that the states in the ``lossy'' subspace, the $\ket{\uparrow}|j\rangle$ states with specific momentum values $p = 2j\hbar k_\text{B} + 2\hbar k_\text{RB}$, could be strongly coupled to a near continuum of states (many, many different final momentum values after multiple absorption-spontaneous emission cycling events), such that the probability of returning to the initial state was essentially zero. Second, we assumed that states from a stable subspace could be weakly coupled to this lossy subspace, at a rate smaller than the loss rate from the unstable states, such that population would not coherently build up in these states, but would instead be effectively lost from the system.

We note that the above description does not involve any explicit particle loss, and could be fully captured by a description involving a continuum of momentum states and the two internal states, $\ket{\downarrow}$ and $\ket{\uparrow}$.
However, true dissipation does in fact enter through the loss of phase coherence during spontaneous emission (considering the information loss to the emitted light fields to be irreversible).
Here, we ask whether genuine dissipation is actually necessary to engineer an effective form of loss in synthetic lattices, or whether weak coupling to a large, empty reservoir of states is sufficient~\cite{Giusteri-LossNoLoss}.
We find evidence for the latter, at least in terms of providing an effective dissipation over some timescale set by the size of the engineered reservoir.
We show, both in simulation and in experiment, how a tunable and local effective loss can be engineered into synthetic lattices even without any true form of dissipation.

For simplicity, we consider the case of a two-site synthetic lattice, a double well with sites $\ket{\text{L}}$ and $\ket{\text{R}}$ coherently coupled with an inter-well tunneling rate $t_\text{sys}$.
By forming this double well from two sites of a larger 1D array, the left and right wells may be coupled, separately, to auxiliary sets of lattice sites which can form the large reservoir of initially unoccupied states.
Here, we restrict ourselves to the scenario in which only the right well experiences coupling to additional states at a rate $t_\text{link}$.
While there is only a single link from $\ket{\text{R}}$ to the left boundary of the reservoir, we can create a situation in which $\ket{\text{R}}$ is effectively irreversibly (on some timescale) coupled to a near continuum of states.
We consider that the reservoir consists of $N \gg 1$ sites, with a large nearest-neighbor coupling rate $t_\text{res}$ that is greater than $t_\text{link}$.
Considering only the reservoir, it will feature a band of delocalized eigenstates with a small energy spacing $\squig 4 t_\text{sys}/N$.
For sufficiently large $t_\text{link}$, the right well simultaneously couples to many unoccupied states of the reservoir, whose time-dependent superposition represents a wavepacket that propagates away from the interface and into the reservoir.
For a sufficiently large reservoir, the revival time of this superposition state, relating to the time it takes to reflect from the right end of the reservoir and return to the system-reservoir interface, can be longer than the time of relevant system dynamics.

\begin{figure*}[t]
	\includegraphics[width=\textwidth]{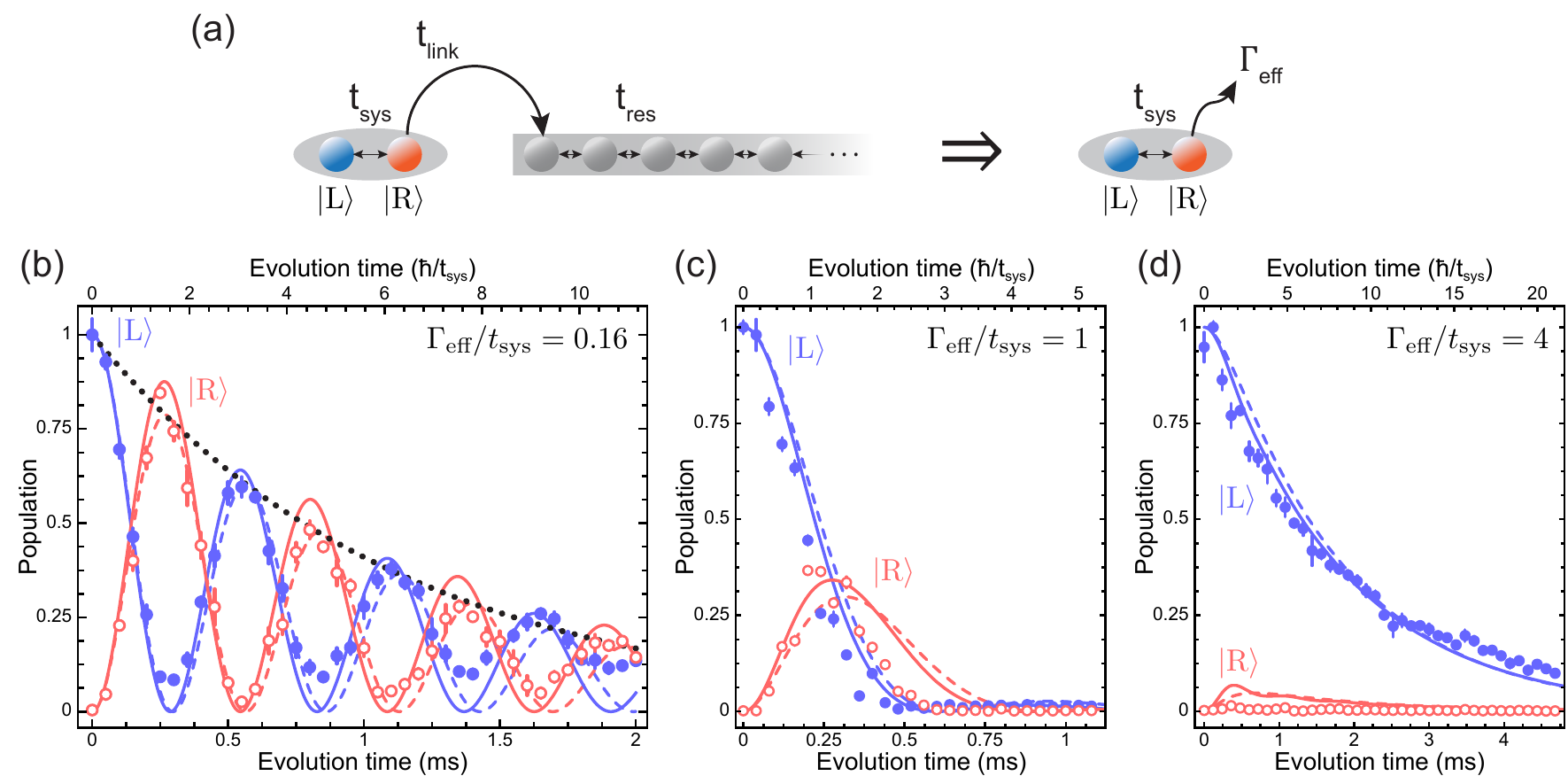}
	\caption{\label{FIG:fig3}
		\textbf{Loss without dissipation.}
		\textbf{(a)}~Implementation of a double well with local loss using a reservoir of unoccupied states. A double well with tunneling $t_\text{sys}$ is coupled via $t_\text{link}$ to a reservoir of states ($N=29$ states in experiment) with tunneling $t_\text{res}$, mimicking an effective loss rate $\Gamma \sim t_\text{link}^2/t_\text{res}$.
		\textbf{(b)}~Population dynamics in $\ket{\text{L}}$ (filled blue circles) and $\ket{\text{R}}$ (open red circles) under small effective loss $\Gamma_\text{eff}/t_\text{sys} = 0.16$. Solid and dashed curves are simulations of the exact two-site model with loss and of the reservoir scheme, respectively, and the dotted black curve shows exponential decay corresponding to $\Gamma_{\text{eff}}/t_{\text{sys}}=0.16$. Simulation curves are plotted with tunnelings $\tslr/h = (1,0.4,1) \times 888$~Hz.
        \textbf{(c)}~Population dynamics under medium effective loss $\Gamma_\text{eff}/t_\text{sys} = 1$. Simulation curves are plotted with tunnelings $\tslr/h = (1,1.25,1.56) \times  612$~Hz.
        \textbf{(d)}~Population dynamics under large effective loss $\Gamma_\text{eff}/t_\text{sys} = 4$. Simulation curves are plotted with tunnelings $\tslr/h = (1,4,4) \times 179$~Hz.
        All error bars denote one standard error of the mean.
	}
\end{figure*}

These dynamics can be viewed purely in terms of the effective loss induced at $\ket{\text{R}}$, with associated loss coefficient $\Gamma_\text{eff} = t^2_\text{link}/t_\text{res}$.
Figure~\pref{FIG:fig3}{a} depicts this mapping between the full reservoir system and the effective model of a double-well system with tunable loss introduced at one well.
We validate this picture of effective loss by comparing numerical simulations of both the full dynamics (with tunneling coefficients $t_\text{sys}$, $t_\text{link}$, and $t_\text{res}$) and the effective dynamics (with equivalent inter-well tunneling coefficient $t_\text{sys}$ and effective right well loss rate $\Gamma_\text{eff} = t^2_\text{link}/t_\text{res}$).
Moreover, we experimentally validate this protocol by realizing a tunable effective loss from one well of a synthetic double well of momentum states.

%As in the first example, we validate this picture of effective loss by comparing numerical simulations of the full dynamics (with tunneling coefficients $t_\text{sys}$, $t_\text{link}$, and $t_\text{res}$) and the effective dynamics (with equivalent intra-well tunneling coefficient $t_\text{sys}$ and effective right-well loss rate $\Gamma_R = t^2_\text{link}/t_\text{res}$). Moreover, we experimentally validate this protocol by realizing a tunable effective loss from one well of a synthetic double well of momentum states.

In both experiment and simulation, we initialize all population in the left well, and monitor the population of both wells over time.
We use a reservoir of $N=29$ lattice sites, sufficiently large such that no population returns to the system from the reservoir.
As shown in Fig.~\pref{FIG:fig3}{b-d}, we investigate three regimes: small effective loss ($\Gamma_\text{eff}/t_\text{sys} = 0.16$), intermediate effective loss ($\Gamma_\text{eff}/t_\text{sys} = 1$), and large effective loss ($\Gamma_\text{eff}/t_\text{sys} = 4$).
We overlay the experimental data with results from numerical simulations of the full system including the reservoir (solid curves) and the effective double well system (dashed curves).

We obtain the experimental tunneling rates by fitting to the data an exact simulation of the momentum-space lattice experiment that accounts for possible off-resonant effects due to the experimental implementation.
This fit procedure finds the appropriate tunneling rates by varying only one free parameter: an overall scaling of the three tunnelings, giving $\tslr/h = (1, 0.4, 1)\times 888$~Hz for the small loss regime, $\tslr/h = (1, 1.25, 1.56)\times 612$~Hz for the intermediate loss regime, and $\tslr/h = (1,4,4)\times 179$~Hz for the large loss regime.
We then use these fitted tunneling rates to generate both displayed simulation curves of the reservoir model and of the effective loss system it mimics.
We note that these tunneling rates roughly match those independently measured through simpler two-site Rabi oscillations: $t_\text{sys}/h = \{976(6), 773(5), 274(1)\}$~Hz for the small, intermediate, and large loss data, respectively.

%These simulation curves are fitted to the data with one free parameter that scales all tunnelings equally, and which agree roughly with our calibrated tunneling rates.\\
%\textbf{HOW THE FITTING PARAMETERS ARE FOUND. eg. The tunneling times used in Fig.~\pref{FIG:fig3} are obtained by performing a residual sum of squares fit between the data and a full simulation of the system based on experimental calibration (+THINGS ALEX AND ERIC KNOW). We obtained tunneling strengths $t_{\text{sys}}/h=$~{888 Hz, 765 Hz, and 716 Hz}, for the small, intermediate and large loss regimes, respectively.}

Under small effective loss (Fig.~\pref{FIG:fig3}{b}), the tunnelling between the wells ($t_{\text{sys}}$) is larger than the transfer out of the system ($t_{\text{link}}$), such that population transfer between the wells dominates over population transfer into the reservoir.
%
%Under small effective loss (Fig.~\pref{FIG:fig3}{b}),  transfer out of the system (represented by $t_{\text{link}}$) is small compared to intra-site tunneling strength ($t_{\text{sys}}$), the rate of population transfer into the reservoir is lower compared to transfer back to the left well.
%
A small fraction of the population is, therefore, lost from the system every full period of oscillation, leading to oscillations that are damped over time.
To highlight this damping, we show an exponential decay curve (dotted black curve) based on the expected loss rate, $\Gamma_{\text{eff}}/t_\text{sys} \approx t_\text{link}^2/t_\text{res} = 0.16$, finding excellent agreement with the data.
In the intermediate regime shown in Fig.~\pref{FIG:fig3}{c}, $t_{\text{link}}$ is slightly larger than $t_{\text{sys}}$. Population briefly builds up in the right well before rapidly tunneling out of the system into the reservoir.
Under large loss (Fig.~\pref{FIG:fig3}{d}), we observe that population transfer out of the left well is vastly reduced compared to the intermediate case. After 5 tunneling times ($5 \hbar/t_\text{sys}$), $\squig 60\%$ of the population still remains in the left well for the large loss case, in contrast to the intermediate loss case where population is entirely within the reservoir. We attribute this difference to the quantum Zeno effect, where strong coupling to a lossy environment actually limits population decay.
We also observe that under strong loss, negligible population builds up in the right well as the inter-well coupling is much smaller than the coupling to the reservoir.
We note that under our chosen tunneling strengths $\tslr/h = (1,4,4) \times 179$~Hz, this scenario is equivalent to a single site with strong loss out of the system.

For all regimes, the simulation of the effective model matches closely with the exact simulation considering an effective loss of the form $\Gamma_\text{eff} = t_\text{link}^2/t_\text{res}$.
Furthermore, the data also shows close agreement with both simulations, with small discrepancies arising due to off-resonant effects in our implementation of the momentum-space lattice~\cite{Meier-AtomOptics}.
We have shown in both experiment and simulation that effective loss can be easily implemented in a non-dissipative system by coupling a small subset of states, representing the system, to a reservoir consisting of the rest of the states.
We further confirm the validity of this scheme by demonstrating the quantum Zeno effect where tunneling out of the left well is reduced for strong loss rates.

Here, for the case of engineering ``loss'' through reversible coupling to a large reservoir of states, we have only explicitly discussed the scenario in which tunable loss appears at the boundaries of a 1D system (specifically, a two-site double well) embedded within a larger 1D lattice.
By simple extension to 2D synthetic lattices~\cite{Gadway-KSPACE,An-FluxLadder}, this approach can also allow for the inclusion of tunable loss at every site of a 1D synthetic lattice.
Generally speaking, extensions to higher dimensions are also possible by embedding the system of choice in an even higher-dimensional system.

\section{Conclusion}

Synthetic lattices allow for engineering Hamiltonians with spectroscopic precision, and have proven to be a versatile platform for exploring the physics of topological and disordered systems.
Here, we have introduced the idea of engineering locally-controlled dissipation in these types of systems, discussing two experimental approaches to introducing site-tunable dissipation.
We have experimentally demonstrated one of the approaches, based on introducing an effective ``loss'' by coupling individual sites to a large, empty reservoir of states, and found good agreement between experiment and the expected non-Hermitian Hamiltonian dynamics.
The introduction of local, tunable loss to the synthetic lattice toolbox should enable future investigations into the interplay of dissipation and topology, disorder, and interactions.
%some extra citations to integrate?~\cite{Li-PT-atoms,Xiao-PT-atoms}.

\section{Acknowledgements}
We thank Kaden R.~A. Hazzard for helpful discussions, and Michael Highman, Sai Paladugu, and Zejun Liu for their careful reading of the manuscript.
This material is based upon work supported by the National Science Foundation under grant No.~1707731.

%\vskip\baselineskip\noindent\textbf{Competing financial interests} The authors declare no competing financial interests.
%
%\vskip\baselineskip\noindent\textbf{Materials and Correspondence} Correspondence and requests for materials should be addressed to B.G.
%
%\vsp\noindent\textbf{Data availability.} All datasets presented here are available from the corresponding author upon request.
\bibliographystyle{apsrev4-1}
%\bibliography{LossSynthBib}
% \printbibliography


\end{document}